\documentclass[a4paper,11pt]{article}
\pdfoutput=1

\usepackage{jcappub}
\usepackage{mathrsfs}
\usepackage{multirow}
\usepackage{mathtools}
\usepackage[dvipsnames]{xcolor}
\usepackage{orcidlink}
\usepackage{tikz}
\usepackage[most]{tcolorbox}
\usepackage{empheq}
\usetikzlibrary{shadows}

\tcbset{
  highlight math style={
    enhanced,
    colframe=blue!40!black,
    colback=blue!10,
    arc=0pt,
    boxrule=1pt,
    drop shadow
  }
}

\title{Perturbations of black holes with primary hair: time evolutions, quasinormal modes and greybody factors}

\author[a,b]{Georgios~Antoniou}
\affiliation[a]{CEICO, Institute of Physics of the Czech Academy of Sciences, Na Slovance 1999/2, 18200, Prague, Czechia}
\affiliation[b]{CENTRA, Departamento de F\'isica, Instituto Superior T\'ecnico - IST, Universidade de Lisboa - UL, Avenida Rovisco Pais 1, 1049 Lisboa, Portugal}
\emailAdd{antoniou@fzu.cz}

\abstract{
In this work we study gravitational perturbations of black holes carrying primary scalar hair in shift-symmetric beyond-Horndeski gravity.
By focusing on the axial sector, we study the geometric structure of the perturbation equations and perform time-domain evolutions both in the physical and in the effective metric descriptions, in order to quantify differences in the response characteristics of solutions with different hair.
In the frequency domain, we then explore the QNM spectrum in the physical-metric formulation, thus verifying and extending existing approximate calculations derived in the effective-metric approach.
We also compute the corresponding greybody factors and absorption cross-sections.
Finally, we use localized deformations of the effective potential as a diagnostic of the relative stability of quasinormal frequencies and greybody-factor observables.
}

\makeatletter
\gdef\@fpheader{}
\makeatother

\begin{document}

\maketitle
\flushbottom
\allowdisplaybreaks

\section{Introduction}

The recent developments in gravitational-wave (GW) astronomy has allowed the field of gravitational physics to enter a new era of observations directly probing the most fundamental aspects of gravity.
In general relativity (GR), the quasinormal-mode (QNM) spectrum of a Kerr black hole is fully determined by its mass and angular momentum.
This property underlies BH spectroscopy: measuring one or more damped oscillation modes provides a way to test whether the remnant is compatible with the Kerr hypothesis~\cite{Kokkotas:1999bd,Berti:2009kk}.
For recent reviews of the theoretical and observational status of black-hole spectroscopy, see Refs.~\cite{Berti:2025spectroscopy,Franchini:2023qnmtests}.
With advanced upcoming detectors on the way, such as LISA and the Einstein Telescope, BH spectroscopy is expected to play a crucial role in testing the strong-field regime of gravity and detecting even small deviations from GR, at unprecedented precision.
The projected fundamental-physics reach of these observatories is summarized in Refs.~\cite{Barausse:2020rsu,Abac:2025sciET}.

There are a number of paths one can follow to modify the Einstein-Hilbert action in pursuit of such deviations.
A theoretically well-motivated arena is provided by scalar-tensor theories, which propagate an additional scalar degree of freedom.
Horndeski gravity and its extensions to beyond-Horndeski and degenerate higher-order scalar-tensor (DHOST) theories constitute such frameworks, involving additional scalars while avoiding Ostrogradsky instabilities~\cite{Horndeski:1974wa,Gleyzes:2014dya,Gleyzes:2014qga,Langlois:2015cwa,BenAchour:2016fzp,Langlois:2017mxy,Kobayashi:2019hrl}.
Shift-symmetric models are especially relevant for BH physics, since they can describe massless scalars which are overall observationally favorable.
Since the action depends on the scalar field only through its derivatives, a scalar profile with linear time dependence,
\begin{equation}
    \varphi(t,r)=qt+\psi(r),
\end{equation}
is compatible with a static and spherically symmetric metric~\cite{Babichev:2013cya,Bakopoulos:2023vkz,Charmousis:2025xug}.
This allows the evasion of some of the assumptions entering standard no-hair theorems while preserving the staticity of the spacetime~\cite{Hui:2012qt}.

Recently, explicit BH solutions with primary scalar hair were constructed in shift- and parity-symmetric beyond-Horndeski/DHOST theories~\cite{Bakopoulos:2023vkz,Bakopoulos:2023hbu}.
The primary nature of the hair is reflected in the fact that the scalar charge is an independent integration constant and thus not fixed by the BH mass, as is usually the case in BHs endowed with secondary hair.
It is associated with the conserved Noether current of the shift symmetry and therefore represents genuine primary hair rather than secondary scalar dressing~\cite{Bakopoulos:2024qll}.
The resulting geometries correspond to modifications of the GR solution, where the latter is approached in the limit where the scalar charge is taken to be zero~\cite{Bakopoulos:2023vkz,Charmousis:2025xug}.
They can therefore serve as natural laboratories for studying how primary hair affects BH perturbations and the corresponding GW observables.
We should also mention that such primary-hair configurations have recently been extended to neutron stars \cite{Boumaza:2026dix}.

Studying the perturbations of such solutions is important if one wants to assess their viability/stability, and/or use them as observational probes of beyond-GR effects in gravitational signatures.
The radial perturbations of primary-hair solutions have recently been investigated in \cite{Charmousis:2026bhe}.
For non-radial perturbations in spherical symmetry the axial sector is significantly easier to explore, since the scalar and gravitational perturbations decouple ~\cite{Regge:1957td,Zerilli:1970se,Kobayashi:2012kh,Langlois:2021aji,Charmousis:2025xug}.
More generally, odd-parity analyses of hairy Horndeski/DHOST black holes have shown that the gravitational sector can possess nontrivial stability conditions~\cite{Ogawa:2015pea,Takahashi:2016dnv,Takahashi:2019oxz,deRham:2019ctd,Tomikawa:2021pca}.
For the primary-hair solutions considered here, the axial perturbations propagate on an effective metric which differs from the physical one seen by matter and electromagnetic fields.
In this effective-metric formulation it has been shown that the axial perturbation problem can be cast into a Schr\"odinger-like equation on the auxiliary geometry ~\cite{Langlois:2022eta,Noui:2023abc}.
In this case therefore, the characteristic surface is determined by the effective metric governing the axial perturbations.
This is important in the context of the implementation of the QNM boundary conditions which must be done on the horizon relevant to the axial perturbations.
Previous studies of these solutions have exploited the effective-metric description to estimate the axial QNM spectrum using WKB methods~\cite{Charmousis:2025xug}.
WKB methods however are approximate and are most reliable for single-barrier potentials and sufficiently large multipole numbers~\cite{Konoplya:2011qq,Matyjasek:2017psv,Konoplya:2026cus}.
A direct integration of the axial system is therefore needed in order to check the WKB estimates, especially for lower multipoles.
In one of the four main parts of this paper, we work with the coupled first-order system of differential equations by employing a direct integration approach in the frequency domain.
This is expected to increase the accuracy of the results since it is not based on the local form of the effective potential near the peak (WKB), but instead solves the full radial boundary-value problem.

In addition, we perform time-domain evolutions in both the physical and the effective metric descriptions.
In the first case we numerically evolve the system of partial differential equations under initial conditions that correspond to a Gaussian pulse, while in the latter, we work with the single axial gravitational degree of freedom in the Regge-Wheeler form under similar initial conditions.
This way we are able to directly compare the waveforms produced via these two methods.

On the other hand, GBFs are real-frequency scattering observables which characterize the probability of transmission through the effective potential barrier and similarly to the QNMs can encode important information about the underlying theory~\cite{Antoniou:2025gbf}.
In practice, their calculation is done through the same system of equations by enforcing boundary conditions which are however different from those corresponding to QNMs.
In the context of BHs with primary hair the scattering problem has recently been studied for test fields~\cite{Antoniou:2025gbf,Bolokhov:2026massive}, but to our knowledge no calculations have been done for the  gravitational sector.
In the present work, the GBFs are obtained by solving the effective-metric master equation with scattering boundary conditions, rather than by using a WKB approximation.
This defines one of the other main goals of this work, namely the use of GBFs as beyond-GR diagnostics for the axial response to perturbations.

The motivation for this comparison is closely connected to the recent literature on the spectral sensitivity of QNM poles and the stability of the corresponding time-domain response~\cite{Jaramillo:2020tuu,Cheung:2021bol,Berti:2022xfj}.
Finally, GBFs have recently attracted renewed interest for ringdown physics because they define observables which have been shown to be more stable to small deviations in comparison with QNMs~\cite{Rosato:2024gbfstability,Oshita:2023bje,Okabayashi:2024gbfii,Rosato:2025freqdomain,Rosato:2026greyring}.
This stability has not been explicitly explored in the context of hairy BHs.
Such a study should provide additional quantitative and qualitative data regarding the relative robustness of QNMs and GBFs under localized deformations.

The paper is organized as follows: in Sec.~\ref{sec:theory} we introduce the class of shift-symmetric DHOST theories considered in this work, review the mathematical formulas describing the solutions with primary scalar hair, and derive the axial perturbation equations in the physical-metric formulation.
We then analyze the characteristic structure of the system of perturbations and perform time-domain evolutions in both the physical and effective-metric descriptions in Sec.~\ref{sec:time_domain}.
In Sec.~\ref{sec:frequency_domain} we study the frequency-domain problem, formulate the direct integration numerical prescription and compute QNMs, GBFs and the corresponding axial absorption cross sections.
Finally, in Sec.~\ref{sec:localized_deformations} we use localized potential deformations to compare the sensitivity of QNM frequencies and GBFs.
We conclude and provide future perspectives in Sec.~\ref{sec:conclusions}.

\section{Theoretical framework}
\label{sec:theory}

\subsection{Background black holes with primary scalar hair}

In this subsection we will revisit the background solutions characterized by primary scalar hair which were first discussed in ~\cite{Bakopoulos:2023vkz}.
These BHs belong to the shift-symmetric subclass of beyond-Horndeski theories.
We can write the action describing this framework as
\begin{equation}
    S=\int d^4x \sqrt{-g}\left[P(X)+F(X)R+\sum_{i=1}^{5}A_i(X)L_i^{(2)}\right],
\label{eq:action}
\end{equation}
where the scalar kinetic term is
\begin{equation}
    X\equiv-\frac12 \nabla_\mu\varphi\nabla^\mu\varphi .
\end{equation}
The restriction to the beyond-Horndeski subclass is ensured via the choices
\begin{equation}
    A_2=-A_1\;,\;\;A_4=-A_3=\frac{A_1+F_X}{X}\;,\;\;A_5=0\, .
\end{equation}
Moreover, homogeneous solutions can be found when \cite{Charmousis:2025xug},
\begin{equation}
    1-F(X)-2X A_1(X)=0.
\end{equation}
These BHs are then characterized by a spacetime metric that can be written as
\begin{equation}
    ds^2=-A(r)dt^2+\frac{dr^2}{A(r)}+r^2d\Omega^2 .
\end{equation}
By respecting the shift symmetry, the scalar field is allowed to have a linear time dependence
\begin{equation}
    \varphi(t,r)=qt+\psi(r),
\end{equation}
where the constant $q$ controls the amount of the scalar charge and therefore the scalar profile.
Then BHs with primary hair can be found for the following choices of the functions appearing in \eqref{eq:action}
\begin{align}
    P(X)
    &=
    -\frac{2\alpha}{\lambda^2}X^p,
    \\
    F(X)
    &=
    1-2X A_1(X),
    \\
    A_1(X)
    &=
    \frac{\alpha}{2}X^{p-1},
    \\
    A_3(X)
    &=
    \frac{\alpha}{2}(2p-1)X^{p-2},
\end{align}
Here $p$ is a positive (half-)integer, $\alpha$ is dimensionless, and $\lambda$ is a length scale.
In this scenario, the kinetic term for the scalar field is given by
\begin{equation}
    X(r)=\frac{q^2}{2}\frac{1}{1+(r/\lambda)^2},
\end{equation}
and the radial derivative of the scalar is expressed as
\begin{equation}
    \psi'(r)^2=\frac{q^2}{A(r)^2}\left[1-\frac{A(r)}{1+(r/\lambda)^2}\right].
\end{equation}
The spacetime metric function can be expressed analytically as
\begin{equation}
    A(r)=1-\frac{2\mu}{r}-\frac{2\lambda\xi_p}{r}\Xi_p\!\left(\frac{r}{\lambda}\right),
\end{equation}
where
\begin{equation}
    \xi_p \equiv \alpha(2p-1) \left(\frac{q^2}{2}\right)^p
\end{equation}
measures the strength of the scalar hair by quantifying the deviation from GR, and
\begin{equation}
    \Xi_p(x) \equiv \int_0^x \frac{u^2 du}{(1+u^2)^p} = \frac{x^3}{3} \,{}_2F_1\!\left(\frac32,p;\frac52;-x^2\right).
\end{equation}
The integration constant $\mu$ is not the usual ADM mass, since there also exists a constant asymptotic contribution from $\Xi_p$.
The Schwarzschild limit is approached by setting $q=0$, or equivalently $\xi_p=0$.

In the rest of this work we specialize to the $p=2$ branch.
It was pointed out in \cite{Charmousis:2025xug} that the Noether charge associated with the primary scalar hair is well defined for $p>3/2$, meaning that $p=2$ is the smallest possible choice for which convergence is ensured.
Moreover, the $p=2$ case leads to a simple closed-form solution that is asymptotically flat while deviating from GR at smaller radii.
We furthermore fix $M=2\lambda=2$ and vary the parameter $\xi\equiv\xi_2$. The metric function for this branch is now
\begin{equation}
    A(r)=1-\frac{2M}{r}+\xi\left[\frac{1}{1+r^2}+\frac{\pi/2-\arctan r}{r}\right].
\end{equation}
This solution is asymptotically flat and reduces continuously to the Schwarzschild result by taking $\xi\to 0$.
The physical event horizon is then defined by
\begin{equation}
    A(r_h)=0 .
\label{eq:physical_horizon}
\end{equation}

\subsection{Axial perturbations}

We will now consider perturbations of the spacetime metric in the axial sector.
Here, the scalar perturbations decouple and we are simply left with the gravitational ones.
The latter ones are schematically given by
\begin{equation}
    g_{\mu\nu}\rightarrow g_{\mu\nu}+\delta g_{\mu\nu},
\end{equation}
and can be written in the Regge-Wheeler gauge~\cite{Regge:1957td,Zerilli:1970se} as
\begin{equation}
    \delta g_{\mu\nu}^{\rm axial}=
    \begin{pmatrix}
    0 & 0 & -h_0(t,r)\,\csc\theta\,\partial_\varphi & h_0(t,r)\,\sin\theta\,\partial_\theta\\
    0 & 0 & -h_1(t,r)\,\csc\theta\,\partial_\varphi & h_1(t,r)\,\sin\theta\,\partial_\theta\\
    * & * & 0 & 0 \\
    * & * & 0 & 0
    \end{pmatrix}
    Y_{\ell m}.
\end{equation}
where we suppress the $(\ell,m)$ labels on $h_0$ and $h_1$ for better visual clarity.
The dynamics are expressed through the two metric perturbation functions $h_0$ and $h_1$.
In the axial case, the nontrivial equations correspond to the $(t,\varphi),\, (r\varphi)$ and $\theta\varphi$ components of the perturbed Einstein equations.
The ones relevant for our analysis can be written schematically as
\begin{align}
    \begin{split}
    &\ddot{h}_0+\alpha_0(r)h_0+\alpha_1(r)h_1+ \alpha_2(r)\dot{h}_0+\alpha_3(r)\dot{h}_1+\alpha_4(r)\dot{h}_0'+\alpha_5(r)\dot{h}_1'=0\, ,
    \end{split}
\label{eq:physical_1}
    \\
    &\ddot h_1 -\dot h_0'+\frac{2}{r}\dot h_0 +\beta_0(r)h_0+\beta_1(r)h_1 =0 \, ,
\label{eq:physical_2}
\end{align}
where overdots denote $\partial_t$, primes denote $\partial_r$, and the radial coefficient functions $\alpha_i(r)$ and $\beta_i(r)$ depend on the background solution and on  $(\ell,\xi)$.
We present the full expressions of the coefficients in the Appendix.
In addition to Eqs. \eqref{eq:physical_1}-\eqref{eq:physical_2} we need to consider the constraint equation
\begin{equation}
\begin{split}
    &\dot{h}_0+\gamma_0(r)\,{h}_0+\gamma_1(r)\,{h}_1+\gamma_2(r)\,\dot{h}_1+\gamma_3(r)\,{h}_0'+\gamma_4(r)\,{h}_1'=0\, .
\end{split}
\label{eq:constraint}
\end{equation}
The detailed expressions for the coefficients appearing in Eqs. \eqref{eq:physical_1}-\eqref{eq:constraint} are given in App. \ref{app:coefficients}.
At the same time, it has been shown that one can study the axial perturbation problem of such theories in an effective metric formulation \cite{Langlois:2022eta,Noui:2023abc,Charmousis:2025xug}.
In this case the axial system can be cast into a single second-order Schr\"odinger-type equation.
Specifically, for the homogeneous $p=2$ branch considered here, and using the notation of Ref.~\cite{Charmousis:2025xug}, the effective line element can be written as
\begin{equation}
    d\hat{s}^2=\sqrt{F(r)}\left[-\Phi(r)d\bar{t}^2+\frac{F(r)}{\Phi(r)}dr^2+r^2d\Omega^2\right].
\label{eq:effective_metric}
\end{equation}
The time coordinate $\bar{t}$ differs from the physical time coordinate $t$ by a radial shift that removes the mixed $dt\,dr$ term in the effective geometry \cite{Charmousis:2025xug}, while the functions entering ~\eqref{eq:effective_metric} are given by
\begin{align}
    \Phi(r)&=A(r)-\frac{\xi}{3(1+r^2)},
    \\
    F(r)&=1-\frac{\xi}{3(1+r^2)^2}.
\end{align}
Here $F(r)\equiv F(X(r))$ is the coupling function $F(X)$ evaluated on the $p=2$ background.
It should not be confused with an independent metric coefficient.
Then, the axial perturbations propagating with respect to this effective metric see the \textit{axial gravitational horizon} which is found by
\begin{equation}
    \Phi(r_g)=0 .
\label{eq:gravitational_horizon}
\end{equation}
Notice that in principle $r_h$ determined from \eqref{eq:physical_horizon} and $r_g$ need not coincide.
In the Schwarzschild limit which is recovered by setting $\xi\to 0$, one has $F(r)\to1$ and $\Phi(r)\to A(r)$, and therefore the two metrics and the two horizons coincide with each other and with the GR values.
We may then define the effective tortoise coordinate as
\begin{equation}
    \frac{d\bar r_*}{dr}= \frac{\sqrt{F(r)}}{\Phi(r)} .
\label{eq:effective_tortoise}
\end{equation}
The master equation for the master variable $\Psi_\ell(\bar t,\bar r_*)$ then takes the usual Schr\"odinger form
\begin{equation}
    \left[-\frac{\partial^2}{\partial \bar t^{\,2}}+\frac{\partial^2}{\partial \bar r_*^{\,2}}-\bar V_\ell(r)\right]\Psi_\ell(\bar t,\bar r_*)=0 \, ,
\label{eq:effective_master}
\end{equation}
where the axial effective potential can be written as
\begin{equation}
    \bar V_\ell(r)=\frac{\ell^2+\ell-2}{r^2}\Phi-\frac{Q}{r}\Phi\Phi'+\frac{2P}{r^2}\Phi^2 ,
\label{eq:effective_potential_p2}
\end{equation}
with the prime denoting differentiation with respect to $r$, and
\begin{align}
    Q&=\frac{rF'+4F}{4F^2},
    \\
    P&=\frac{r^2\left(7F'^2-4FF''\right)+16rFF'+32F^2}{32F^3}\, .
\end{align}

\section{Time-domain evolutions}
\label{sec:time_domain}

In this section we will study in detail the characteristic structure of the system of equations describing the perturbations \eqref{eq:physical_1}-\eqref{eq:physical_2}, deduce the propagation speeds and recover the gravitational horizon from the effective metric description.
We will also perform calculations in the time domain by solving both the system \eqref{eq:physical_1}-\eqref{eq:physical_2} in the physical metric formulation, and the single master equation \eqref{eq:effective_master} in the effective-metric approach.

\subsection{Principal symbol and axial characteristic speeds}

Here we will explore the mathematical structure of the system of equations \eqref{eq:physical_1}-\eqref{eq:physical_2}.
We will extract the characteristic speeds and deduce the horizon relevant for the propagation of the axial perturbations.
For convenience, we may collect the perturbation variables into the field vector
\begin{equation}
\mathbf{h}
=
\begin{pmatrix}
h_0\\ h_1
\end{pmatrix}.
\end{equation}
The principal part of the equations can then be written in matrix notation as
\begin{equation}
    E_I^{\rm prin}=\mathbf{K}_{ij}\,\partial_t^2 h^j+\mathbf{M}_{ij}\,\partial_t\partial_r h^j+\mathbf{G}_{ij}\,\partial_r^2 h^j,
\end{equation}
where $i,j=1,2$.
For the axial system under consideration the matrices appearing above are given by
\begin{equation}
    \mathbf{K}=
    \begin{pmatrix}
    1 & 0\\
    0 & 1
    \end{pmatrix}
    \;,\;
    \mathbf{G}=
    \begin{pmatrix}
    0 & 0\\
    0 & 0
    \end{pmatrix}\;,
    \;
    \mathbf{M}=
    \begin{pmatrix}
    a(r) & b(r)\\
    -1 & 0
    \end{pmatrix}.
\label{eq:KMG_matrices}
\end{equation}
In terms of the schematic coefficients in Eqs.~\eqref{eq:physical_1}-\eqref{eq:physical_2}, one may identify $a(r)=\alpha_4(r)$ and $b(r)=\alpha_5(r)$, after the same normalization of the two equations used in Eq.~\eqref{eq:KMG_matrices}.
It is straightforward to see from the explicit expression of $\alpha_5$ provided in App. \ref{app:coefficients}, that for the $p=2$ backgrounds, the condition $b(r)=0$ is equivalent to $\Phi(r)=0$, and therefore reproduces the axial gravitational horizon defined through Eq.~\eqref{eq:gravitational_horizon}.

If we introduce the covector $\zeta_\mu dx^\mu=\zeta_tdt+\zeta_rdr$, we can obtain the principal symbol by replacing $\partial_\mu\to\zeta_\mu$ in the highest-derivative part of the system ~\eqref{eq:physical_1}-\eqref{eq:physical_2}, so that we get $\mathcal{P}(\zeta)=\mathbf{K}\zeta_t^2+ \mathbf{M}\zeta_t\zeta_r+\mathbf{G}\zeta_r^2$.
In our case this yields
\begin{equation}
    \mathcal{P}(\zeta)
    =
    \begin{pmatrix}
    \zeta_t^2+a(r)\zeta_t\zeta_r & b(r)\zeta_t\zeta_r\\
    -\zeta_t\zeta_r & \zeta_t^2
    \end{pmatrix}.
\label{eq:Pmatrix}
\end{equation}
From \eqref{eq:Pmatrix} we can determine the characteristic covectors by demanding that $\det\mathcal{P}(\zeta)=0$.
If we also define the radial characteristic speed by $c=-\zeta_t/\zeta_r$, we find that the two propagating axial characteristic speeds are given by
\begin{equation}
    c_\pm(r) =\frac{a(r)\pm\sqrt{\Delta_c(r)}}{2}\; , \; \Delta_c(r)=a(r)^2-4b(r).
\end{equation}
The sign of the quantity $\Delta_c(r)$ determines whether the system is hyperbolic ($\Delta_c(r)>0$), elliptic ($\Delta_c(r)<0$), and parabolic ($\Delta_c(r)=0$).
The characteristic constant-$r$ surfaces correspond to $b(r_g)=0$, which is equivalent to the axial-horizon condition \eqref{eq:gravitational_horizon}.
For $p=2$ hyperbolicity is ensured for $3(r^2+1)^2>\xi$.
For all numerical cases we consider in the following sections, we restrict $\xi\le 2.5$, to ensure this condition is satisfied, while allowing us to obtain significant deviations from GR.

\subsection{Time-domain evolution in the physical metric}

\begin{figure}
    \centering
    \includegraphics[width=\linewidth]{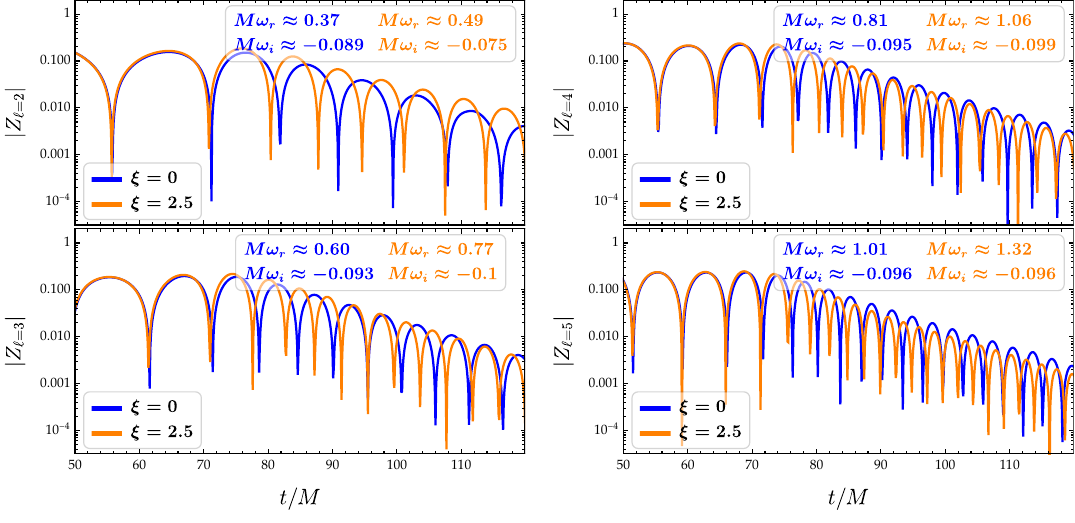}
    \caption{Time-domain integration of the axial system in the physical-metric formulation for an initial Gaussian pulse. Each panel shows the diagnostic waveform $Z(t,r)$ for $\ell=[2,3,4,5]$ as a function of the normalized time. We extract the waveform at $r_{\rm obs}=50M$.}
    \label{fig:time_physical}
\end{figure}

We now choose to evolve the axial system \eqref{eq:physical_1}-\eqref{eq:physical_2} directly in the physical-metric variables.
The inner edge of the numerical domain of integration is chosen just outside the axial gravitational horizon, $r_{\rm min}=r_g+\epsilon$, with $\epsilon$ taken sufficiently small.
In practice we set $\epsilon=10^{-5}M$ and checked that our results do not change for smaller values.
To tackle the problem numerically, we first reduce the second-order system to first order in time by defining the auxiliary field vector $\boldsymbol{P}$
\begin{equation}
    \boldsymbol{P}=
    \begin{pmatrix}
    p_0\\ p_1
    \end{pmatrix}
    \equiv
    \begin{pmatrix}
    \partial_t h_0\\ \partial_t h_1
    \end{pmatrix},
\end{equation}
so that the system can be written schematically as
\begin{align}
    \partial_t\boldsymbol{h}&=\boldsymbol{P},
\label{eq:systemEq1}
    \\
    \partial_t\boldsymbol{P}&=\boldsymbol{f}(\boldsymbol{h},\boldsymbol{P},\partial_r\boldsymbol{P}).
\label{eq:systemEq2}
\end{align}
To ensure numerical stability, the system is evolved in the physical tortoise coordinate $x$, defined by
\begin{equation}
    \frac{dx}{dr}=\frac{1}{A(r)} .
\end{equation}
The characteristic combinations associated with the principal part are found to be
\begin{equation}
    W_\pm=p_0+\left(a-c_\pm\right)p_1.
\end{equation}
At the left $x=x_L$ and right $x=x_R$ boundaries, we impose the corresponding dissipative conditions
\begin{align}
    W_{{\rm in},L}&=p_0(t,x_L)+\left[a_L-c^{(x)}_{{\rm in},L}\right]p_1(t,x_L)=0,
    \\
    W_{{\rm in},R}&=p_0(t,x_R)+\left[a_R-c^{(x)}_{{\rm in},R}\right]p_1(t,x_R)=0.
\end{align}
where the subscript in,$i$ with $i=(L,R)$ symbolizes the incoming mode at the respective boundary.
Moreover, $a_{L,R}=a(r(x_{L,R}))$, and $c^{(x)}_{{\rm in},L/R}$ denotes the speed with the appropriate sign characterizing the ingoing mode at each boundary, in the tortoise coordinate, i.e. $c^{(x)}=c/A(r)$.

The diagnostic waveform is taken to be the GR-inspired combination
\begin{equation}
    Z(t,r)=\frac{A(r)}{r}h_1(t,r).
\end{equation}
In terms of initial conditions we specify a Gaussian pulse on the diagnostic function
\begin{equation}
    Z(0,x)=\mathcal{A}\exp\left[-\frac{(x-x(r_0))^2}{2\sigma_x^2}\right]\, ,
\end{equation}
while we set the remaining freely specified initial data to zero, i.e.
\begin{equation}
    h_0(0,x)=0\; \; ,\;\; p_1(0,x)=0.
\end{equation}
Notice however, that the initial value of $p_0$ cannot be chosen freely, but is rather fixed by the axial constraint \eqref{eq:constraint}.

In Fig.~\ref{fig:time_physical} we show some representative waveforms for $\xi=0,2.5$ shown with blue and orange colors respectively, and $\ell=[2,3,4,5]$.
The initial pulse is characterized by $\sigma_x=1.5M$, $\mathcal{A}=1$, and $r_0=6M$.
The waveform is extracted at a fixed observer radius $r_{\rm obs}=50M$.
We see that after the prompt response at early times, the waveform enters the ringdown regime from which we may extract the real and imaginary parts of the frequency.
Specifically, the signal is fitted by the dominant damped sinusoid
\begin{equation}
    Z_{\rm obs}(t)\simeq\mathcal{A}\,e^{\omega_I(t-t_0)}\cos\left[\omega_R(t-t_0)+\varphi_0\right]\, ,
\end{equation}
and the corresponding values are shown in the panels of Fig. \ref{fig:time_physical}.
For $\xi=0$ the GR values are recovered for all multipoles.
For $\xi=2.5$ the frequency of oscillations increases in comparison with the GR value by approximately $30-40\%$ for all multipoles, while the imaginary part remains relatively close to the GR counterpart value except for $\ell=2$ when it decreases in absolute value by around $15\%$.
We stress that while evolving the system \eqref{eq:systemEq1}-\eqref{eq:systemEq2} we monitor the constraint equation \eqref{eq:constraint}, both through its RMS value and its $L^\infty$ norm and verify that it remains small over the interval we use to extract the QNM frequencies.

\subsection{Time-domain evolution in the effective metric}

\begin{figure}
    \centering
    \includegraphics[width=\linewidth]{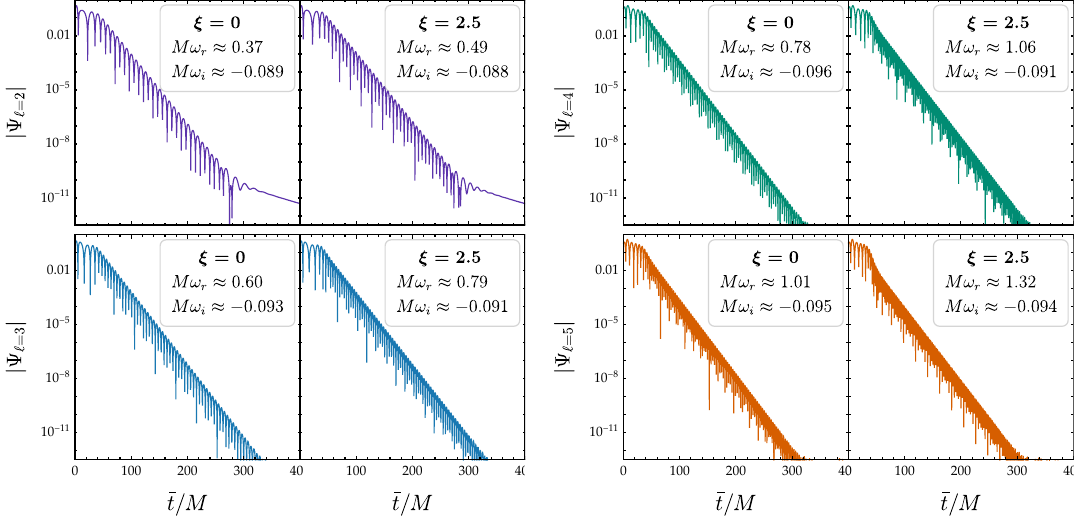}
    \caption{Time-domain evolution of the axial master variable in the effective geometry formulation. We extract the waveform at distance $r_{\rm obs}=25M$. For each multipole number $\ell=\{2,3,4,5\}$, we show the GR waveform corresponding to $\xi=0$ and the hairy waveform with $\xi=2.5$. We also display the fitted real and imaginary parts of the dominant mode in each case. We use the same horizontal and vertical ranges to allow for direct visual comparison.}
    \label{fig:eff_time}
\end{figure}

We will now study the effective master variable whose evolution is described by Eq. \eqref{eq:effective_master}.
This approach makes explicit how the axial gravitational horizon and the effective tortoise coordinate enter the dynamical problem.
It will provide a complementary study of the time-domain problem for the same axial gravitational degree of freedom without resorting to the original perturbation variables $(h_0,h_1)$, but rather on the effective metric description developed in \cite{Langlois:2022eta,Noui:2023abc,Charmousis:2025xug}.

Once again we choose to work with a tortoise coordinate, which in this case is the effective tortoise coordinate introduced in \eqref{eq:effective_tortoise}, so that $\bar r_*\in[x_{\rm min},x_{\rm max}]$, with the inner edge placed close to the axial gravitational horizon and the outer edge in the asymptotic region.
The potential \eqref{eq:effective_potential_p2} is then evaluated as a function of $\bar r_*$ by interpolation using the radial map $\bar r_*(r)$.
To study the time evolution we excite the system with a localized Gaussian pulse
\begin{equation}
    \Psi(0,x)=A_0\exp\left[-\frac{(x-x_0)^2}{2\sigma^2}\right]\; ,\;\partial_{\bar{t}}\Psi(0,x)=0,
\end{equation}
where in this case $x\equiv\bar r_*$.

Near the axial gravitational horizon, $\bar V_\ell\to 0$, and the ingoing solution behaves as $\Psi\sim e^{-i\omega(\bar{t}+x)}$.
We therefore impose
\begin{equation}
    \left.\left(\partial_{\bar{t}}-\partial_x\right)\Psi\right|_{x=x_{\rm min}}=0 \,.
\end{equation}
At spatial infinity the outgoing solution behaves as $\Psi\sim e^{-i\omega(\bar{t}-x)}$, so that we have to impose
\begin{equation}
    \left.\left(\partial_{\bar{t}}+ \partial_x\right)\Psi\right|_{x=x_{\rm max}}=0 \, .
\end{equation}
We then extract the waveform at a fixed observer position which is taken to be $r_{\rm obs}=25M$.

We show the results for $\ell=\{2,3,4,5\}$ and $\xi=\{0,2.5\}$ in Fig. \ref{fig:eff_time}.
We once again see that after the initial time corresponding to the prompt response, the waveform enters the ringdown phase from which we may extract the real and imaginary parts of the mode similarly to what we did in the previous subsection, through a damped sinusoidal fit.
Even though in this work we are interested in the ringdown phase, we can see that in the $\ell=2$ case, for $t/M\gtrsim 300$ the late-time tail behaviour appears.
We chose to display the same vertical and horizontal range in all scenarios, so as to allow for visual comparison of the waveforms.
Compared to the results obtained via the physical metric formulation followed in the previous subsection, we notice greater stability at later times which allows us to evolve the waveforms for longer periods, ensuring that we lie well within the ringdown regime.
The extracted real parts of the dominant modes are in good agreement with respect to the ones derived earlier.
In terms of the imaginary parts discrepancies arise which we attribute to the fact that for the physical metric results, the fit was performed in a shorter temporal window, before numerical artifacts emerge.
This can affect the sinusoidal fit significantly rendering it less trustworthy.
Specifically, in Fig. \ref{fig:time_physical} we see that $t_{\rm max}/M \sim 120$ while from Fig. \ref{fig:eff_time} we notice the ringdown range to span from $t/M\sim 80$ all the way to $t/M\sim 260$ for all multipoles, thus allowing for a much more accurate fit.
The subsequent rigorous analysis in the frequency domain in Sec. \ref{sec:frequency_domain} demonstrates the excellent quantitative agreement between the two formulations, providing a strong numerical consistency check between the two approaches.
Overall, however, we deduce that the trends observed in Fig. \ref{fig:time_physical} and Fig. \ref{fig:eff_time} are similar: increasing the hair parameter $\xi$ leads to significantly faster oscillations for all $\ell$ while the effect on the damping time is less significant.

\section{Frequency-domain analysis and observables}
\label{sec:frequency_domain}

In this section we will work in the frequency domain and derive both the QNM spectrum and GBFs associated with the hairy BH solutions.
To this end, we decompose the metric perturbation functions as
\begin{align}
    h_{(0,1)}(t,r)=& \int d\omega \; h_{(0,1)}(r) \, e^{-i \omega t}\, ,
    \label{eq:Fourier_1}\\
    \Psi(\bar{t},r)=& \int d\omega \; \Psi(r) \, e^{-i \omega \bar{t}}\, .
    \label{eq:Fourier_2}
\end{align}

Like in the time-domain analysis, we may study the frequency-domain problem both in the physical and effective metric descriptions.
Below we present the corresponding mathematical formulations.
In particular, for the QNM numerical analysis in subsection \ref{subsec:QNMs} we employ both the physical and effective-metric formulations, allowing a direct comparison between them.
We also employ the master effective description to specifically apply the WKB method as an independent benchmark.
In Sec. \ref{subsec:GBFs}, for the scattering problem, we only work with the effective-metric master equation, for which the transmission and reflection amplitudes are defined in the usual way (see \eqref{eq:infPsi}-\eqref{eq:GBF}).

After Fourier decomposing according to \eqref{eq:Fourier_1}, the linearized equations \eqref{eq:physical_1}-\eqref{eq:physical_2} reduce to a radial first-order system of the form
\begin{equation}
    \frac{d}{dr}
    \begin{pmatrix}
    h_0\\ h_1
    \end{pmatrix}
    =
    \mathbf{M}(r,\omega;\ell,M,\xi)
    \begin{pmatrix}
    h_0\\ h_1
    \end{pmatrix}\, ,
    \label{eq:first_order_h0h1}
\end{equation}
where the $2\times 2$ matrix $\mathbf{M}$ depends on the background solution.
In order to apply a direct integration approach later on, it is important to derive the appropriate boundary conditions at the axial gravitational horizon and at asymptotic infinity.
Near the axial horizon $r_g$, we write the associated Frobenius expansions as
\begin{align}
    h_0(r)&=(r-r_g)^{\rho_{0}^{\rm in/out}}\sum_{n=0}^{\infty} a_n^{\rm in/out}(r-r_g)^n\, ,
    \\
    h_1(r)
    &=(r-r_g)^{\rho_{1}^{\rm in/out}}\sum_{n=0}^{\infty} b_n^{\rm in/out}(r-r_g)^{n-1}\, .
\end{align}
The exponents $\rho_i^{\rm in/out}$ and the relative coefficients $b_n/a_n$ are fixed order by order by the physical-metric perturbation equations.
In the Schwarzschild limit, where $r_g=r_h$, these expansions reduce to the standard Regge-Wheeler ingoing and outgoing series.
However we should clarify a subtle but important point: in order to retrieve the GR limit, one has to set $\xi=0$ in \eqref{eq:first_order_h0h1} before substituting the near-horizon expansions.
In other words, we found that the GR limit, $\xi\to 0$ cannot be approached smoothly in the physical metric formulation after expanding around $r_g$.
Rather, if one wants to derive the GR indicial structure, one has to consider the $\xi\to 0$ at the level of the perturbation equations, before taking the Frobenius expansion.
Since the latter is performed around the $\xi$-dependent point $r_g$, the resulting indicial roots and coefficients contain terms that become degenerate if we take the limit $\xi \to 0$ subsequently.
This nonuniformity concerns the near-horizon representation of the physical variables and does not prevent the physical solutions, including the QNM frequencies, from approaching their GR counterparts as $\xi\to 0$ in a continuous manner.

At spatial infinity, the system \eqref{eq:first_order_h0h1} admits the usual pair of oscillatory solutions corresponding to ingoing and outgoing waves,
\begin{align}
    h_0(r)=&\; e^{\pm i\omega r}r^{\pm 2iM\omega}\sum_{n=0}^{\infty}\frac{A_n^{\rm in/out}}{r^{n-1}}\, ,
    \\
    h_1(r)
    =&\; e^{\pm i\omega r}r^{\pm 2iM\omega}\sum_{n=0}^{\infty}\frac{B_n^{\rm in/out}}{r^{n-1}}\, ,
\end{align}
where the $+$ and $-$ signs correspond to outgoing and ingoing solutions respectively.

Regarding the effective metric description in the frequency domain, after applying the decomposition \eqref{eq:Fourier_2} the master equation \eqref{eq:effective_master} describing the axial dynamics takes the Schr\"odinger-like form
\begin{equation}
    \frac{d^2\Psi}{d\bar r_*^2}+\left[\omega^2-\bar V_\ell(r)\right]\Psi=0\, .
\end{equation}
Now, the near-axial-horizon and infinity expansions for the QNMs/scattering problem are written as
\begin{align}
    \Psi =&\; e^{\mp i\omega \bar r_*} \sum_{n=0}^\infty a_n^{\rm in/out}(r-r_g)^n \;\;,\;\; \bar r_*\rightarrow -\infty,\\
    \Psi =&\; e^{\pm i\omega \bar r_*} \sum_{n=0}^\infty \frac{a_n^{\rm in/out}}{r^n} \;\;,\;\; \bar r_*\rightarrow \infty \, .
\end{align}

\subsection{Quasinormal modes}
\label{subsec:QNMs}

\begin{figure}
    \centering
    \includegraphics[width=\linewidth]{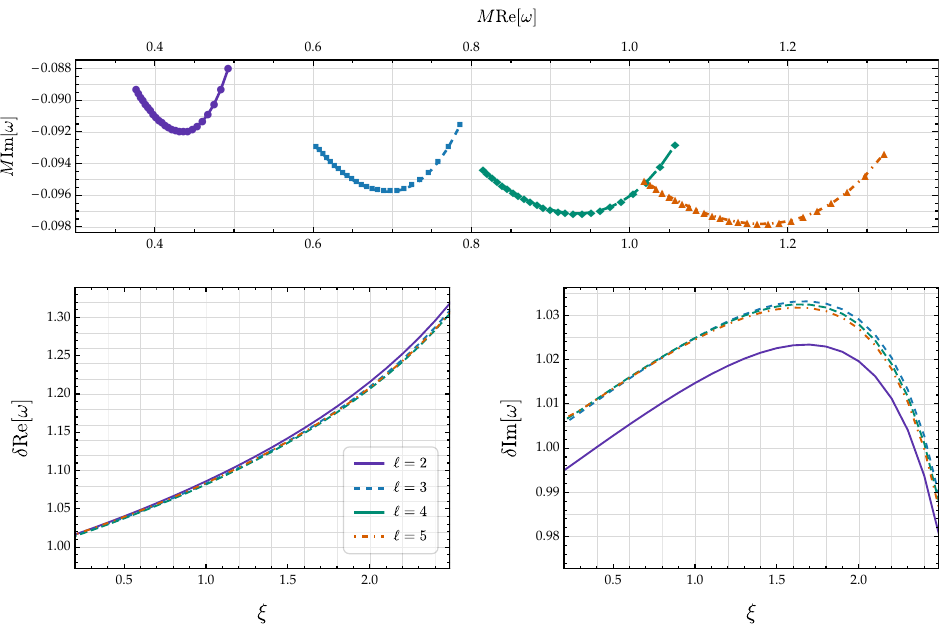}
    \caption{Top: real and imaginary parts of the fundamental axial QNM for $\ell=2,3,4,5$ as functions of the scalar-hair parameter $\xi\in[0.1,2.5]$, with spacing $\delta\xi=0.1$. Bottom: relative difference of the real and imaginary parts of the QNM with respect to their GR counterparts as functions of $\xi$.}
    \label{fig:QNMs}
\end{figure}

\begin{figure}
    \centering
    \includegraphics[width=\linewidth]{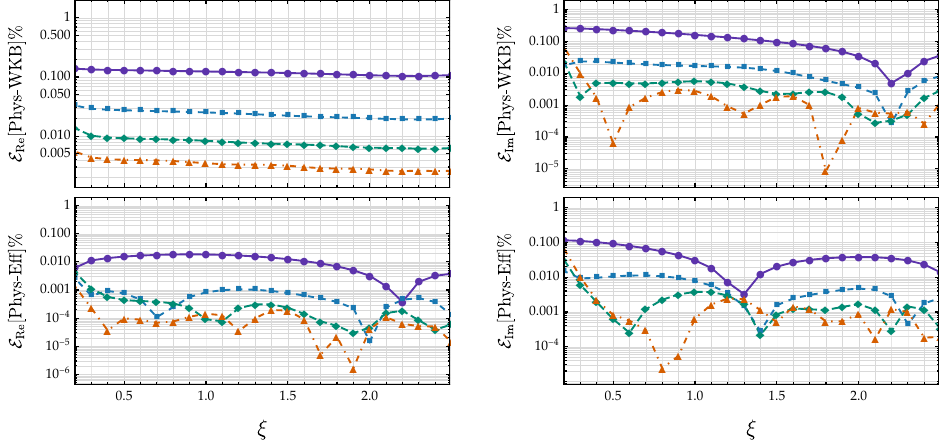}
    \caption{Top: relative difference between the QNM frequencies obtained by DI of the physical-metric system and by the 6th-order WKB method applied to the effective-metric potential.
    Bottom: relative difference of the results derived in the physical and effective-metric descriptions.}
    \label{fig:QNMs_errors}
\end{figure}

QNMs correspond to solutions of the axial system \eqref{eq:first_order_h0h1} satisfying purely ingoing boundary conditions at the axial gravitational horizon and purely outgoing ones at spatial infinity.
In this work we employ both the physical and the effective metric approach in solving for QNMs.
In both cases we perform two integrations, one from the axial horizon outwards (in practice we start very close to it, i.e. at $r_g+2\times 10^{-8}M$) and one from infinity inward.
The maximum expansion order at the two boundaries is truncated so that the numerical results converge with increased accuracy.
For the physical metric problem, we construct the $2\times2$ matrix made of the two independent solutions of Eq.~\eqref{eq:first_order_h0h1}:
\begin{equation}
    {\textbf X}_{\rm phys}
    = \begin{bmatrix}
    h_0^{(r_g)} & h_0^{(\infty)}\\
    h_1^{(r_g)} & h_1^{(\infty)}\\
 \end{bmatrix}\, .
\end{equation}
On the other hand, in the effective-metric description the matrix we construct is made of the solutions to the homogeneous equation and their first derivatives, i.e.
\begin{equation}
    {\textbf X}_{\rm eff}
    = \begin{bmatrix}
    \Psi^{(r_g)} & \Psi^{(\infty)}\\
    \partial_r \Psi^{(r_g)} & \partial_r \Psi^{(\infty)}\\
 \end{bmatrix}\, .
\end{equation}

Then, QNM frequencies correspond to roots of the determinant of the matrix ${\textbf X}_{\rm phys}$, \textit{i.e.} they are given by solving:
\begin{equation}
    \textnormal{det}\ \textbf{X}(\omega)|_{r_m}=0\, ,
\end{equation}
at some intermediate matching radius $r_m$.
In our numerical solutions we always ensure that the shooting method result is independent of $r_m$.
The solutions provide a discrete set of complex frequencies,
\begin{equation}
    \omega=\omega_{\ell n} =\omega_R+i\omega_I,
\end{equation}
where $n$ is the overtone number.
The real part $\omega_R$ determines the oscillation frequency and the imaginary one $|\omega_I|^{-1}$ gives the damping time.

Fig. \ref{fig:QNMs} groups the results we derived for the QNMs.
The top panel shows the results obtained by DI in the frequency domain, using the effective-metric description for multipole numbers $\ell=\{2,3,4,5\}$.
The vertical axis shows the imaginary part of the QNM associated with the damping time and the horizontal one the oscillation frequency.
Each distinct point corresponds to a mode with $\xi$ in the range $[0.1,2.5]$ with $\delta \xi=0.1$.
In the small-$\xi$ limit of the parameter space, the GR value is smoothly approached.
For any $\ell$, the maximum deviation of the real part of the QNM is encountered at the max value of $\xi$ considered, namely $\xi=2.5$, while the imaginary part presents a nonmonotonic behaviour.
It is worth pointing out that degeneracies may appear for larger multipoles, as it becomes evident by the intersection of the $\ell=4$ and $\ell=5$ lines.
In the second row of Fig. \ref{fig:QNMs} we show the relative differences of the real and imaginary parts of the QNM with respect to GR, as a function of the parameter $\xi$, to further elucidate the explicit dependence.

To quantify the potential deviations that may arise if we follow the two different descriptions, or if we employ different numerical techniques we also present Fig. \ref{fig:QNMs_errors}. The top row shows the relative error between the results obtained with DI in the physical metric formulation and 6th-order WKB in the effective-metric description.
The deviations are of the order of $0.5\%$ for $\ell=2$ and decrease for higher multipoles both for the real and imaginary parts.
The second row shows relative differences for the modes derived by DI in physical and effective metric formulations.
The trend is overall similar to the top row, but quantitatively the differences are much smaller, demonstrating good agreement between the two descriptions.

\subsection{Greybody factors and absorption cross section}
\label{subsec:GBFs}

\begin{figure}
    \centering
    \includegraphics[width=\linewidth]{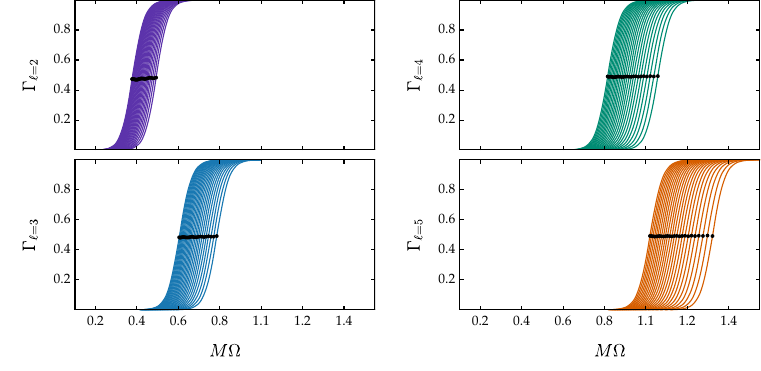}
    \caption{Gravitational greybody factors in the axial sector for the hairy black holes with $\ell=\{2,3,4,5\}$. The hair parameter is taken in the range $\xi\in[0.1,2.5]$, with spacing $\delta\xi=0.1$. The black points denote the real part of the corresponding QNM frequency.}
    \label{fig:GBFs}
\end{figure}

To study the scattering problem we need to consider boundary conditions describing purely ingoing waves at the horizon and ingoing-outgoing waves at infinity.
The QNM results from the previous subsection suggest that both the physical and the effective metric descriptions yield good agreement when studying the perturbation problem.
Here we will work with the effective master equation \eqref{eq:effective_master} which allows the transmission and reflection amplitudes to be defined directly.
If we normalize the incoming wave at infinity to unity we may express the boundary conditions as
\begin{align}
    \lim_{\bar r_*\rightarrow -\infty}\Psi \sim & \mathcal{A}_\ell(\Omega) e^{-i\Omega \bar{r}_*}\, ,
    \label{eq:infPsi}
    \\
    \lim_{\bar r_*\rightarrow +\infty}\Psi \sim & e^{-i\Omega \bar r_*}+\mathcal{R}_\ell(\Omega)e^{+i\Omega \bar r_*}\, ,
\end{align}
where $\mathcal{A_\ell}(\Omega$) and $\mathcal{R}_\ell(\Omega)$ are the transmission and reflection amplitudes as functions of the scattering frequency.
The greybody factor is then defined as
\begin{equation}
    \Gamma_\ell(\Omega) \equiv\left|\mathcal{A}_\ell(\Omega)\right|^2=1-\left|\mathcal{R}_\ell(\Omega)\right|^2\, ,
\label{eq:GBF}
\end{equation}
where the latter equality follows from the conservation of the radial flux.
Therefore, since GBFs probe a different type of aspect of the response compared to QNMs, namely they describe real-frequency transmission through the potential barrier, they provide a strong, complementary observable.

In Fig. \ref{fig:GBFs} we plot the GBFs for multipole numbers $\ell=\{2,3,4,5\}$ as functions of the scattering frequency.
For each multipole the hair-parameter $\xi$ ranges from $0.1$ to $2.5$ (similarly to what we showed in Fig. \ref{fig:QNMs} for the QNMs), with step $\delta\xi=0.1$.
We see that for all multipoles, increasing $\xi$ results in the GBF shifting to higher frequencies, while the rate of the displacement increases when increasing $\ell$, a behaviour that is consistent with the one encountered in the QNM analysis.
With black dots we also denote the real part of the corresponding fundamental QNM frequency $\Re [\omega_{0\ell}(\xi)]$.
We see a very close agreement of the latter with the scattering frequency $\Omega_{1/2}$ corresponding to the point where the GBF is one half, i.e. $\Gamma(\Omega_{1/2})=1/2$.
Finally, the degeneracy observed between $\ell=4$ and $\ell=5$ QNMs is also established here, as is observed in the right panel column of Fig. \ref{fig:GBFs}.

\begin{figure}
    \centering
    \includegraphics[width=0.55\linewidth]{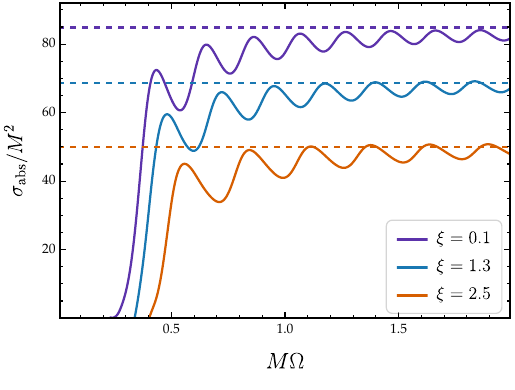}
    \caption{Axial-sector absorption cross section constructed from the gravitational greybody factors and normalized by $M^2$. The horizontal reference value denotes the geometric-optics limit $\sigma_{\rm geo}^{\rm ax}=\pi b_c^2$, where $b_c$ is the critical impact parameter associated with the axial effective potential.}
    \label{fig:absorption_cross_sections}
\end{figure}

Moreover, the GBFs can be combined into a dimensionful absorption cross section which can be much more informative than specific GBFs.
In a spherically symmetric scattering problem, the partial absorption cross section for a specific multipole $\ell$ is defined as
\begin{equation}
    \sigma_{\ell}(\Omega)=\frac{\pi}{\Omega^2}(2\ell+1)\Gamma_{\ell}(\Omega)\, ,
\end{equation}
where the factor $(2\ell+1)$ ensures the consideration of degenerate modes with different azimuthal numbers in a spherically symmetric background.
To find the axial absorption cross section we have to sum with respect to the multipole ~\cite{Dolan:2008kf}, i.e.
\begin{equation}
    \sigma(\Omega)=\sum_{\ell}\sigma_{\ell}(\Omega).
    \label{eq:total_absorption_general}
\end{equation}
Since in this work we explore the axial sector in Fig. \ref{fig:absorption_cross_sections} we show the axial absorption cross section normalized with the square of the BH mass, for three representative cases of hairy solutions with $\xi=\{0.1,1.3,2.5\}$.
To calculate it we perform the summation in \eqref{eq:total_absorption_general} from $\ell_\text{min}=2$.
In practice, we found that it is sufficient to truncate the numerical calculation at $\ell_\text{max}=10$.

The low-frequency behaviour of $\sigma$ depends on the spin and on the lowest allowed multipole.
Since we consider gravitational perturbations starting at $\ell=2$, the axial absorption cross section vanishes in the low-frequency limit.
Indeed, all curves in Fig. \ref{fig:absorption_cross_sections} approach zero in the small-frequency limit.
At high frequency, the partial-wave sum approaches the geometric-optics limit which is related to the axial effective light ring \cite{Decanini:2011xw}.
Specifically, for the axial sector the relevant critical orbit is determined by the leading eikonal part of the axial potential, $\bar V_\ell\sim \ell^2\Phi(r)/r^2$.
The axial light-ring radius $r_c$ therefore should be defined by the condition
\begin{equation}
    \left.\frac{d}{dr}\left(\frac{\Phi(r)}{r^2}\right)\right|_{r=r_c}=0\, .
\end{equation}
Since the associated impact parameter is $b_c^2=r_c^2/\Phi(r_c)$, the high-frequency limit should be
\begin{equation}
    \sigma_{\rm geom}^{\rm ax}=\pi b_c^2=\frac{\pi r_c^2}{\Phi(r_c)} .
\end{equation}
In the GR limit where the hairy parameter vanishes, we recover the usual metric element $\Phi=A=1-2M/r$, and the standard results $r_c=3M$ and $\sigma_{\rm geom}=27\pi M^2$.
In Fig. \ref{fig:absorption_cross_sections} and in the high-energy limit, all curves approach the geometric optics result which is depicted by dashed lines.

\section{Localized deformations and greybody-factor stability}
\label{sec:localized_deformations}

\begin{figure}
    \centering
    \includegraphics[width=\linewidth]{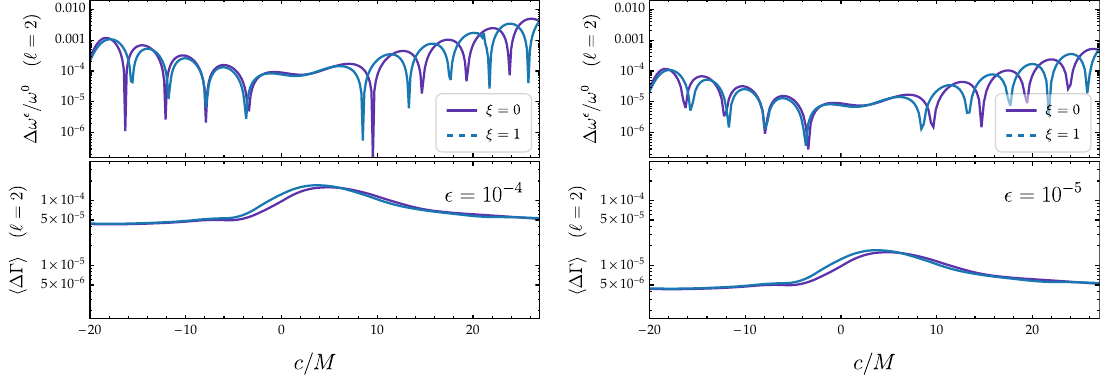}
    \caption{Sensitivity of the fundamental axial QNM frequency and of the integrated greybody factor to a localized P\"oschl-Teller deformation of the effective potential for $\ell=2$.
    The deformation of the potential is centered around the tortoise position $c$.
    The left and right panels correspond to diffent choices of $\epsilon$, namely $\epsilon=10^{-4}$ and $\epsilon=10^{-5}$ respectively.}
    \label{fig:QNM_GBF_stability}
\end{figure}

As a final calculation in this work and in order to further test the robustness of the axial scattering observables, we will consider perturbations of the effective potential by a localized P\"oschl-Teller bump in the tortoise coordinate,
\begin{equation}
\begin{split}
    \bar V_{\ell}(\bar r_*)\longrightarrow\bar V_{\ell}^{\epsilon}(\bar r_*)=& \; \bar V_{\ell}(\bar r_*)+\delta V(\bar r_*),
    \\
    \delta V(\bar r_*)=& \; \frac{\epsilon}{M^2}\operatorname{sech}^2 \left(\frac{\bar r_*-c}{M}\right),
\end{split}
\end{equation}
where $\epsilon\ll1$ controls the amplitude of the deformation and $c$ determines its location.
As mentioned in the introduction, the stability of GBFs under such localized deformations has been explored \cite{Rosato:2024gbfstability}, but to our knowledge no explicit tests have been performed in beyond-GR BH scenarios.
Following \cite{Rosato:2024gbfstability}, we compare the sensitivity of the QNM spectrum and the GBF by defining the following quantities
\begin{equation}
    \Delta_{\rm QNM}(\xi,\epsilon,c)=\left|\frac{\omega_{\ell n}^{\epsilon}(\xi,c)-\omega_{\ell n}^{0}(\xi)}{\omega_{\ell n}^{0}(\xi)}\right|,
\end{equation}
and
\begin{equation}
    {\Delta \Gamma}_{\ell}(\xi,\epsilon,c)=\frac{\int d\Omega\,\left|\Gamma_{\ell}^{\epsilon}(\Omega;\xi,c)-\Gamma_{\ell}^{0}(\Omega;\xi)\right|}{\int d\Omega\,\Gamma_{\ell}^{0}(\Omega;\xi)}.
\end{equation}
In Fig. \ref{fig:QNM_GBF_stability} we plot the diagnostic quantities as a function of the bump location, for $\epsilon=10^{-4}$ and $\epsilon=10^{-5}$ and for two separate scenarios: one for $\xi=0$ which corresponds to a GR solution, and a hairy BH with $\xi=1$.
We notice that when $c$ lies close to the maximum of the potential barrier, both the QNM frequencies and the GBFs respond perturbatively, with the latter displaying its largest deformation.
However, for $|c|\gg M$, the QNM spectrum can become much more sensitive and display an approximately exponential growth, while the integrated GBF deformation remains of order $\epsilon$, as becomes evident by comparing the plots for the two different $\epsilon$ values we considered.
This suggests that GBFs may in principle be better scattering observables compared to QNMs, at least in terms of their stability under small deformations even for BHs characterized by hair.

\section{Conclusions}
\label{sec:conclusions}

In this work, we explored in detail axial perturbations of black holes with primary scalar hair in shift-symmetric beyond-Horndeski/DHOST gravity.
Working directly with the physical-metric perturbation variables, we extracted the characteristic structure of the axial system and identified the gravitational horizon relevant for the propagation of the axial perturbations.
This direct formulation clarifies how the characteristic surfaces of the gravitational perturbations can differ from the physical horizon seen by matter whenever the physical and effective geometries are distinct.

We then performed time-domain evolutions in both formulations studying in particular the ringdown regime.
The physical-metric evolution tests the characteristic boundary prescription in the original variables, while the effective-metric evolution follows the Regge-Wheeler-type master equation.
By fitting the waveform we extracted the real and imaginary part of the dominant QNM.
The time-domain study provides a novel diagnostic and a useful independent check of the frequency-domain spectrum which followed.

In the frequency domain, we computed the QNM spectrum by directly integrating both the physical-metric system and the master equation, without resorting to approximate methods, and explored many different multipoles.
We compared the results obtained with WKB estimates obtained from the effective-metric potential, and demonstrated that the agreement is overall good (less than 1\%) and improving for higher multipoles.
This comparison shows explicitly how WKB estimates derived in previous works \cite{Charmousis:2025xug} compare against the direct boundary-value problem.
We also confirmed that the results between the frequency and time-domain analyses are compatible.

Subsequently, we studied the scattering problem and computed the GBFs by solving the effective-metric system with the appropriate boundary conditions.
The GBFs complement the QNM calculation by probing the transmission properties of the same axial wave operator rather than its complex-frequency poles.
We also derived the axial absorption cross section, which makes the low and high-frequency regimes easier to interpret.
In particular, the high-frequency limit is controlled by the geometric-optics cross section associated with the axial effective light ring. The low-frequency behaviour on the other hand, reflects the fact that the radiative gravitational sector starts at $\ell=2$.

Finally, we considered localized deformations of the effective potential in order to diagnose the relative robustness of QNMs and GBFs.
We saw that these can induce large shifts in QNMs while leaving integrated GBF observables perturbatively small.
This suggests that the use of GBFs as stable scattering observables for hairy black holes beyond GR can be advantageous compared to QNMs, and motivates further work on their role in ringdown tests of modified gravity.

Finally, we should once again emphasize that our main frequency-domain results did not resort to approximate approaches and were also performed in the physical formulation where the definition of an effective metric is not required.
Thus, a solid framework has been established for the study of polar perturbations as a natural extension of the current work.

\section*{Acknowledgements}

The author conducts their research as part of the ``RINGMOD'' project, that is part of the Marie Sk{\l}odowska-Curie Actions - COFUND project ``P4F'' which is co-funded by the European Union, Physics for Future Grant Agreement No. \href{https://cordis.europa.eu/project/id/101081515}{101081515}.
The author acknowledges financial support provided by FCT - Funda\c{c}\~ao para a Ci\^encia e a Tecnologia, I.P., through the ERC-Portugal program Project ``GravNewFields''. The author also thanks the Funda\c{c}\~ao para a Ci\^encia e a Tecnologia (FCT), Portugal, for the financial support to the Center for Astrophysics and Gravitation (CENTRA/IST/ULisboa) through grant No.~\href{https://doi.org/10.54499/UID/PRR/00099/2025}{UID/PRR/00099/2025} and grant No.~\href{https://doi.org/10.54499/UID/00099/2025}{UID/00099/2025}.

\appendix

\section{Coefficients}
\label{app:coefficients}
Here we provide the expressions for the coefficients $\alpha_i(r)$, $\beta_i(r)$ and $\gamma_i(r)$ appearing in Eqs.~\eqref{eq:physical_1}-\eqref{eq:constraint}.
\begin{align}
    \mathcal{R}(r)&\equiv r^2+\lambda^2\, ,\\
    \mathcal{S}(r)&\equiv \sqrt{\mathcal{R}(r)-\lambda^2 A(r)}\, ,\\
    \mathcal{L}&\equiv \ell(\ell+1)-2\, ,\\
    D(r)&\equiv \lambda^2\mathcal{R}(r)\xi+\left[3\mathcal{R}(r)^2-\lambda^4\xi\right]A(r)\, .
\end{align}
We also define the auxiliary functions
\begin{align}
    \begin{split}
        P_1(r)\equiv& -\mathcal{L}\lambda^2\mathcal{R}(r)^2\xi+3A(r)\big[\mathcal{L}\mathcal{R}(r)^3-4r^2(r^2-\lambda^2)\lambda^2\xi+r\mathcal{R}(r)^3(2A'(r)\\
        &+rA''(r))\big]\, ,
    \end{split}
    \\
    \begin{split}
        P_2(r)\equiv& -2\left(5r^2\lambda^2+2\lambda^4\right)A(r)^2+2r\mathcal{R}(r)^2A'(r)\\
        &+\mathcal{R}(r)A(r)\left[8r^2+4\lambda^2-r\lambda^2A'(r)\right]\, ,
    \end{split}
    \\
    \begin{split}
        P_3(r)\equiv & +6r\lambda^2A(r)^2-2\mathcal{R}(r)^2A'(r)=-\mathcal{R}(r)A(r)\left[4r-\lambda^2A'(r)\right]\, .
    \end{split}
\end{align}
The coefficients entering Eq. \eqref{eq:physical_1} are
\begin{align}
    \alpha_0(r)&=\frac{\mathcal{L}\lambda^4\xi^2\mathcal{S}(r)^2}{3r^2\mathcal{R}(r)D(r)}\, ,\\
    \alpha_1(r)&=\frac{\lambda^2\xi A(r)\mathcal{S}(r)P_1(r)}{3r^2\mathcal{R}(r)^{5/2}D(r)}\, ,\\
    \alpha_2(r)&=\frac{\lambda^2\xi P_2(r)}{2r\sqrt{\mathcal{R}(r)}\mathcal{S}(r)D(r)}\, ,\\
    \alpha_3(r)&=-\frac{A(r)^2\left[2r\lambda^2\xi+3\mathcal{R}(r)^2A'(r)\right]}{D(r)}\, ,\\
    \alpha_4(r)&=-\frac{2\lambda^2\sqrt{\mathcal{R}(r)}\xi A(r)\mathcal{S}(r)}{D(r)}\, ,\\
    \alpha_5(r)&=\frac{\mathcal{R}(r)A(r)^2\left[\lambda^2\xi-3\mathcal{R}(r)A(r)\right]}{D(r)}\, .
\end{align}
The coefficients entering Eq. \eqref{eq:physical_2} are
\begin{align}
    \beta_0(r)&=\frac{\mathcal{L}\lambda^2\xi\mathcal{S}(r)}{3r^2\mathcal{R}(r)^{3/2}A(r)}\, ,\\
    \beta_1(r)&=\frac{P_1(r)}{3r^2\mathcal{R}(r)^3}\, .
\end{align}
Finally, the coefficients entering the constraint equation, Eq. \eqref{eq:constraint}, are
\begin{align}
    \gamma_0(r)&=-\frac{\lambda^2\xi P_3(r)}{2\sqrt{\mathcal{R}(r)}\mathcal{S}(r)D(r)}\, ,\\
    \gamma_1(r)&=\alpha_3(r)\, ,\\
    \gamma_2(r)&=\alpha_4(r)/2\, ,\\
    \gamma_3(r)&=\gamma_2(r)\, ,\\
    \gamma_4(r)&=\alpha_5(r)\, .
\end{align}

\bibliographystyle{JHEP}
\bibliography{bibnote}

\end{document}